\documentclass[manuscript]{aastex}
\usepackage{txfonts}
\usepackage{amssymb}
\usepackage{lscape}
\usepackage{graphicx}
\usepackage{color}

\shorttitle{Multi-epoch VLBA observations of 3C 66A} \shortauthors{Zhao et al.}

\begin{document}


\title{Multi-epoch, multi-frequency VLBI study of the parsec-scale jet in the TeV blazar 3C 66A}


\author{Guang-Yao ~Zhao\altaffilmark{1,2},
Yong-Jun ~Chen\altaffilmark{1,3}, Zhi-Qiang ~Shen\altaffilmark{1,3}, Hiroshi ~Sudou\altaffilmark{4}, Satoru ~Iguchi\altaffilmark{5}}


\altaffiltext{1}{Key Laboratory for Research in Galaxies and
Cosmology, Shanghai Astronomical Observatory, Chinese Academy of
Sciences, Shanghai 200030, China}
\altaffiltext{2}{Korea Astronomy and Space Science Institute, Daejeon 305-348, Korea; email: gyzhao@kasi.re.kr}
\altaffiltext{3}{Key Laboratory of Radio Astronomy, Chinese Academy of Sciences, China}
\altaffiltext{4}{Faculty of Engineering, Gifu University, Gifu 501-1193, Japan}
\altaffiltext{5}{National Astronomical Observatory of Japan, Tokyo 181-8588, Japan}


\begin{abstract}
We present the observational results of the $\gamma$-ray blazar, 3C 66A, at 2.3, 8.4, and 22 GHz at 4 epochs during 2004-05 with the VLBA. The resulting images show an overall core-jet structure extending roughly to the south with two intermediate breaks occurring in the region near the core. By model-fitting to the visibility data, the northmost component, which is also the brightest, is identified as the core according to its relatively flat spectrum and its compactness. As combined with some previous results to investigate the proper motions of the jet components, it is found the kinematics of 3C 66A is quite complicated with components of inward and outward, subluminal and superluminal motions all detected in the radio structure. The superluminal motions indicate strong Doppler boosting exists in the jet. The apparent inward motions of the innermost components last for at least 10 years and could not be caused by new-born components. The possible reason could be non-stationarity of the core due to opacity change.
\end{abstract}



\section{Introduction}           
\label{sect:intro}

The radio source, 3C 66A (also known as B0219+428, NRAO102, 4C
42.07), is a well-known blazar, which is further sub-classified as an
intermediate synchrotron peaked BL Lac object (IBL) ( Abdo et al.
2010). Since its optical counterpart was identified by
Wills \& Wills (1974), 3C 66A was studied over a wide frequency
range, from radio to $\gamma$-ray. It is also one of the sources detected in TeV (Aliu et al. 2009, Acciari et al. 2009). Like other blazars, 3C 66A
exhibits violent variabilities at radio, IR, optical, and X-ray (e.g.
B\"{o}ttcher et al. 2005 hereafter B05), and in recent
years, flares at $\gamma$-ray wavelengths have also been detected (e.g.
Swordy 2008, Abdo et al.  2011).

On kpc-scales, 3C 66A shows a 6 arcsecond extended structure at
the position angle (P.A.) $\sim$ 170$\degr$ (Price et al.
 1993) with two weak lobes, one at $15^{''}$ south of the
central region, the other at $10^{''}$ with P.A. about $-20\degr$ (
Taylor et al. 1996). Very Long Baseline Interferometry
(VLBI) images from 2.3 to 43 GHz show a typical core-jet structure
on parsec scales (e.g. Taylor er al. 1996, Jorstad et
al. 2001, Marscher et al. 2002, Cai et
al. 2007, hereafter C07). The kinematics of 3C 66A jet has
been studied via multi-epoch VLBI observations with superluminal
motions detected for several jet components (e.g. Jorstad et
al. 2005, Britzen et al. 2008).

 The redshift value of 3C 66A is adopted to be z=0.444(Miller et al.~\cite{mill78}) by several previous works (e.g. B05 and C07) . This value was measured based on the detection of only one single line and is argued to be quite uncertain (e.g. Bramel et al. 2005). Recently, several redshift ranges for this source were obtained by different methods, e.g. $z < 0.58$ by Yang \& Wang (2010), $z > 0.30$ by Shaw et al. (2013), $0.335 < z < 0.41$ at 99 percent confidence and $0.335 < z < 0.444$ at 99.9 percent confidence by Furniss et al. (2013), and $z > 0.42$ by Stadnik \& Romani (2014). In this work, we adopt $0.335 < z < 0.444$ as the redshift range of 3C 66A.

In this paper we present our multi-frequency, multi-epoch VLBA
observation results of 3C 66A. The observations and data reduction are described in
Sect. 2. The results, including the images, spectra, and proper motion analysis are presented in Sect. 3. In Sect. 4, a
detailed discussion is made with a summary followed in
Sect. 5. In this paper,we use the $\Lambda$CDM cosmological model with $H_0=71$~km~s~$^{-1}$~Mpc$^{-1}$,
$\Omega_M=0.27$, and $\Omega_\Lambda=0.73$ (Spergel et al. 2003).

\section{Observations and data reduction}
\label{sect:Obs}

3C 66A was observed as phase-reference calibrator for the VLBA
projects BS087 and BS144, which aimed at searching for evidence of a
super-massive black hole binary in the nearby radio galaxy 3C 66B
(Sudou et al. 2003, Zhao et al. 2011). These
projects were carried out during 2001-2002 (Session 1) and 2004-2005
(Session 2), respectively. Session 1 includes 6 observations at 2.3 and 8.4
GHz, and 2 observations at 22 GHz at the first 2 epochs. The results of Session 1 were published in C07. During Session 2, 3C 66A was successfully observed at 2.3, 8.4, and 22 GHz at all 4
epochs, 2004.80, 2005.05, 2005.35, and 2005.54. The total on-source
time of 3C 66A at each epoch is about 80 min at 2.3 GHz, 30 min at 8.4 GHz, and
40 min at 22 GHz. The data were recorded in VLBA format with 2
intermediate frequency (IF) bands at 2.3 and 8.4 GHz and 4 IF bands
at 22 GHz. Each IF has a bandwidth of 8 MHz. The total recording rate was
128 Mbps using a 2-bit sampling mode. The observed data were
correlated with the VLBA correlator in Socorro.

The output data were then loaded into the NRAO AIPS package for further
processing. Amplitude calibration was performed by using the measured system
temperature and antenna gain curves. Phase calibrations were made in a
standard way given in AIPS COOKBOOK, except that additional correction for all
of the observations in Session 2 was made to correct the possibly inaccurate
Earth orientation parameters (EOPs) used by the VLBA correlator from 2004 to 2005.
The residual phase
delays and delay rates that were mainly due to antenna position
errors and atmospheric phase fluctuations were calibrated using
global fringe-fitting (Schwab \& Cotton 1983).
Bandpass calibration was performed for higher sensitivity using 0133+476
for all epochs. The output data from AIPS were
then read into the Caltech DIFMAP package for hybrid mapping and
model fitting. Circular Gaussian models were used in model fitting, which
may reduce the number of free parameters
and give more consistent centroid positions across the epochs
(Lister et al. 2009).
The corresponding errors for the fitted parameters were estimated by using the
expressions described in Lee et al. ~\cite{lee08}, except the flux
density errors were taken to be 10\% at 2.3 and 8.4 GHz, and
20\% at 22 GHz, and the errors for the core separations of the
components were taken to be at least 0.8 mas at 2.3 GHz, 0.2 mas at 8.4 GHz,
and 0.06 mas at 22 GHz. Such criterions were adopted based on some previous
works on other radio sources (e.g. C07, Lu et al. 2011).
\begin{table*}[]
  \caption[]{Description of VLBA maps of 3C~66A shown in Fig.~\ref{fig:1} }
  \label{tab:1}
  \begin{tabular}{ccccccc}
\hline\noalign{\smallskip} \hline\noalign{\smallskip}
&&&\multicolumn{3}{c}{Restoring Beam}\\ \cline{4-6}
   $\nu$ & Epoch      & S$_{peak}$      & Major     & Minor     & P.A.& Contours   \\
   (GHz)  & (yr)   & (Jy/beam) &(mas)   &(mas)   &(deg)&(mJy/beam)  \\
   (1)    &(2)        &(3)        &(4)     &(5)     &(6)  &(7)        \\
  \hline\noalign{\smallskip}
2.3 GHz   &2004.80     & 0.453      & 5.86    & 3.77 &-0.74&0.9$\times$(-1,1,2,4,8,16,32,64,128,256,512)  \\
          &2005.05     & 0.476      & 5.89    & 3.58 &-3.21&0.9$\times$(-1,1,2,4,8,16,32,64,128,256,512)  \\
          &2005.35     & 0.424      & 7.51    & 4.25 &-18.5&1.0$\times$(-1,1,2,4,8,16,32,64,128,256)  \\
          &2005.54     & 0.465      & 6.31    & 3.96 &-0.72&0.9$\times$(-1,1,2,4,8,16,32,64,128,256,512)  \\\hline
8.4 GHz   &2004.80     & 0.451      & 1.61    & 1.10 & 5.22&1.2$\times$(-1,1,2,4,8,16,32,64,128,256)  \\
          &2005.05     & 0.466      & 1.51    & 1.01 &0.472&1.0$\times$(-1,1,2,4,8,16,32,64,128,256)  \\
          &2005.35     & 0.496      & 2.11    & 1.18 &-18.2&1.4$\times$(-1,1,2,4,8,16,32,64,128,256)  \\
          &2005.54     & 0.433      & 1.56    & 1.04 &-1.05&1.2$\times$(-1,1,2,4,8,16,32,64,128,256)  \\\hline
22.2 GHz  &2004.80     & 0.464      & 0.829   & 0.575&-0.43&2.0$\times$(-1,1,2,4,8,16,32,64,128)  \\
          &2005.05     & 0.421      & 0.614   & 0.371&-3.29&1.5$\times$(-1,1,2,4,8,16,32,64,128,256)  \\
          &2005.35     & 0.488      & 0.991   & 0.761& 3.72&2.1$\times$(-1,1,2,4,8,16,32,64,128)  \\
          &2005.54     & 0.367      & 0.629   & 0.378&-6.29&1.8$\times$(-1,1,2,4,8,16,32,64,128) \\
  \noalign{\smallskip}\hline\vspace{0.5mm}
  \end{tabular}
\\
Notes:(1) Observing frequency;
      (2) Observing epoch;
      (3) Peak flux density;
      (4), (5), (6) Parameters of the restoring Gaussian beam: the full
width at half maximum (FWHM) of the major and minor axes and the
position angle (P.A.) of the major axis.
      (7) Contour levels of the map. The lowest contour level is three times the rms noise in the
      map.
\end{table*}

\section{Results}
\label{sect:results} 
\begin{figure}
   \centering
   \includegraphics[width=12cm, angle=0]{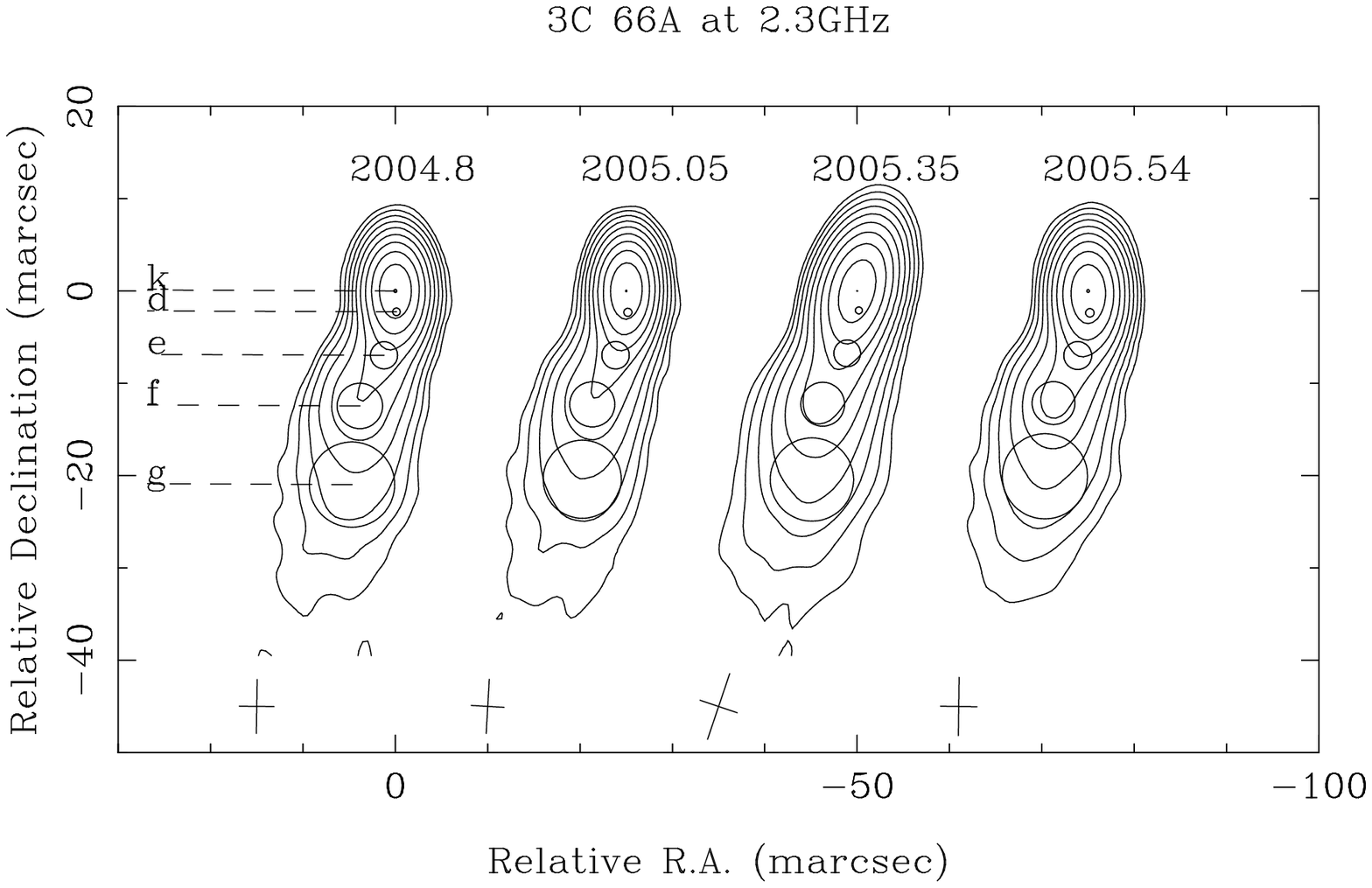}
   \includegraphics[width=12cm, angle=0]{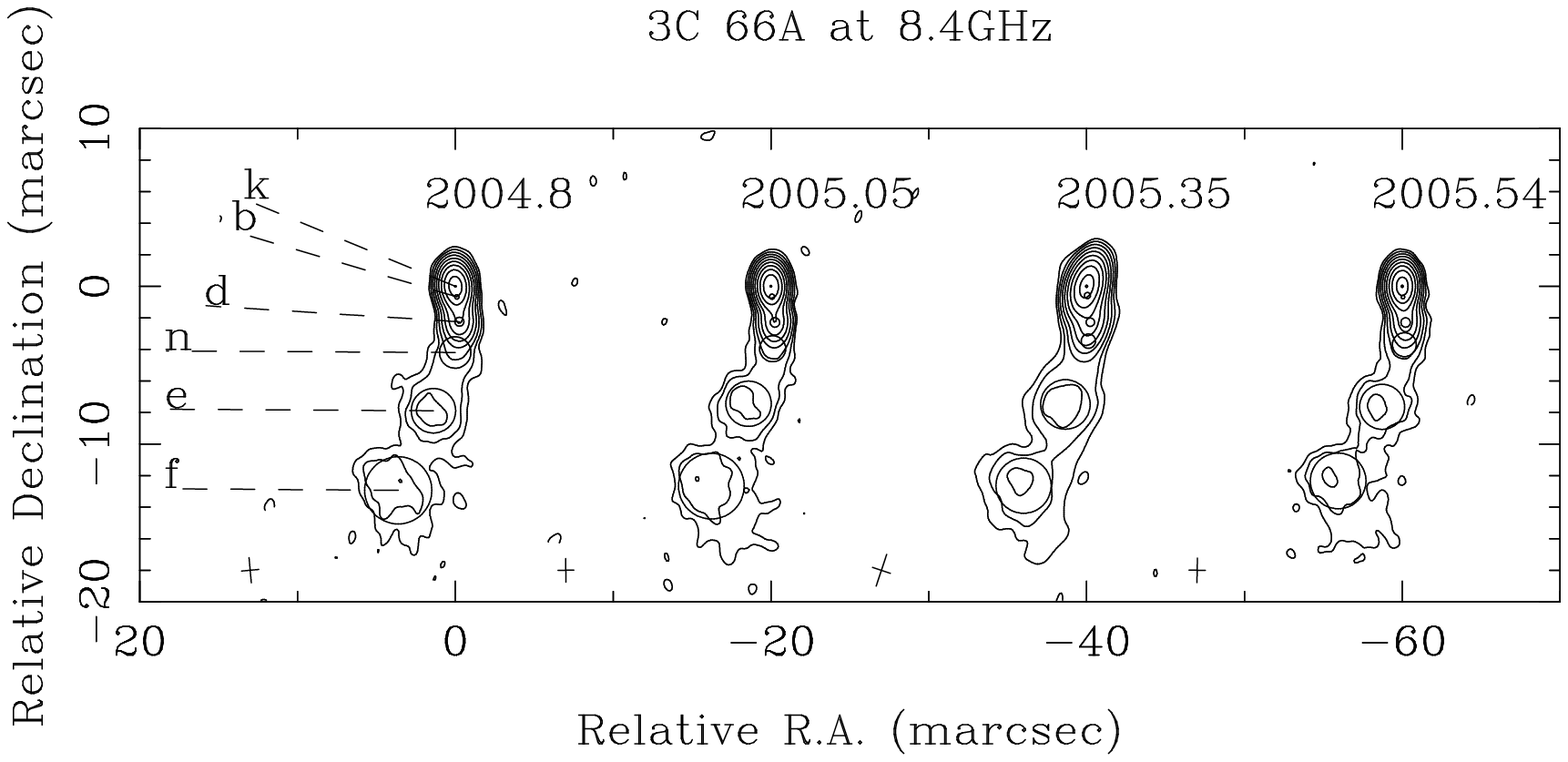}
   \includegraphics[width=12cm, angle=0]{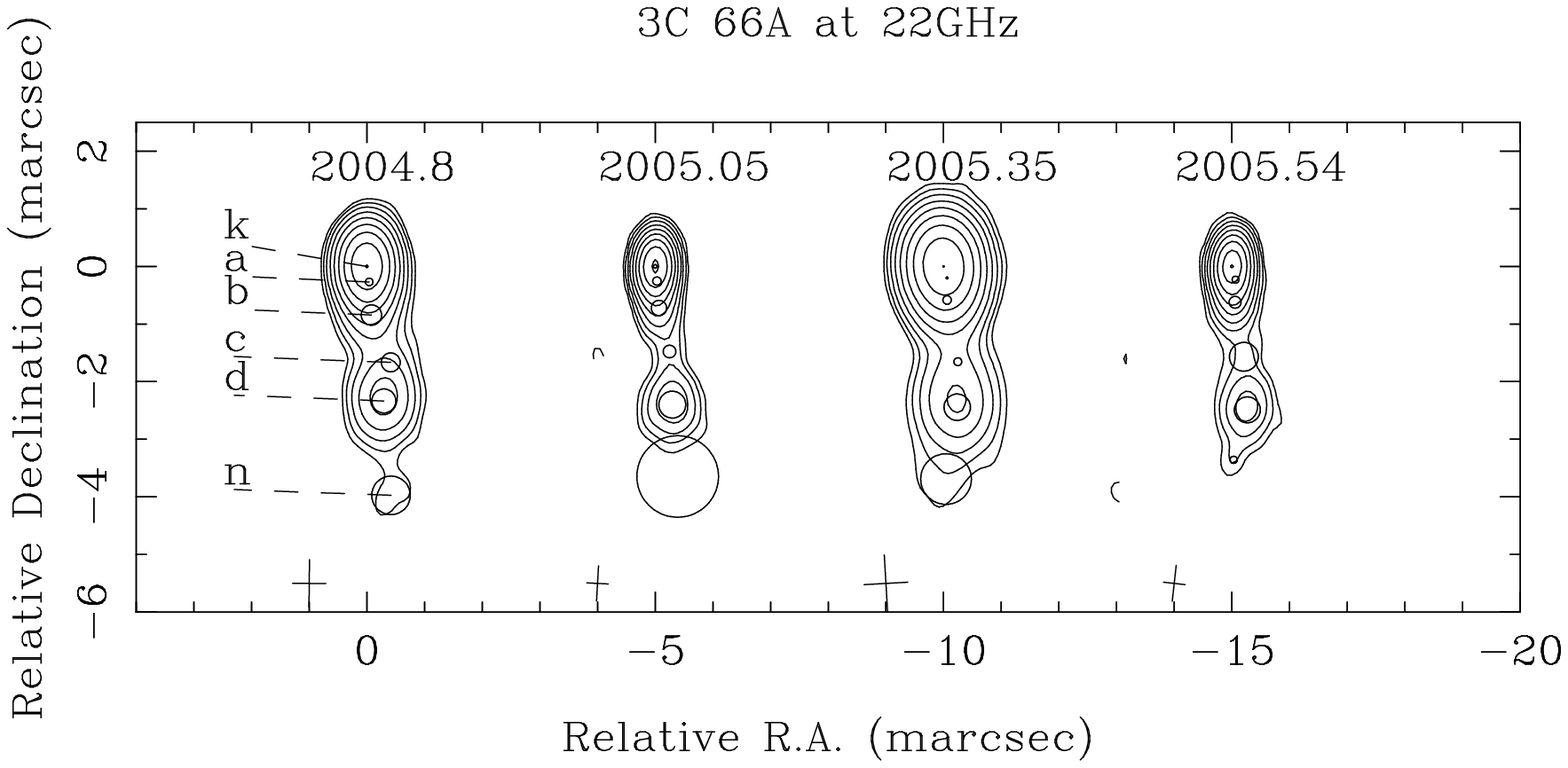}
   \caption{The naturally-weighted VLBA maps of
3C~66A at 2.3 GHz (top), 8.4 GHz (middle), and 22.2 GHz (bottom). The circles
superimposed on the maps represent the Gaussian model components
listed in Table ~\ref{tab:2}, whose names are labeled. The crosses
represent the restoring beams in Table ~\ref{tab:1}. }
   \label{fig:1}
   \end{figure}

In Fig.~\ref{fig:1}, we show the resulting contour
images of 3C 66A at all 4 epochs at 2.3, 8.4, and 22 GHz, with
specific parameters shown in Table.~\ref{tab:1}. An overall core-jet emission
distribution can easily be found at low frequency of 2.3 GHz with higher
dynamic range and larger emission structure, while more specific structures are obtained at 22 GHz with higher resolution. The radio
structure at all these frequencies and epoches shows a consistent
intensity distribution with a compact emission feature lying at the
extreme end of a onesided jet, and others extending to the south.
From the 8.4 GHz radio structures at all 4 epochs, a prominent bending
was found at $\sim4$ mas. And in the 22 GHz images, a minor bending occurs at $\sim1.2$ mas.

To better understand its intensity distribution down the jet, model fitting was done to the remarkable jet features by using the \emph{modelfit} task in DIFMAP. All the resulting components are marked with circular signs with their sizes scaling to their true emission regions. Table.~\ref{tab:2} shows the designations of each jet component, the corresponding flux densities and uncertainties at all epochs and frequencies. The designation of each component mainly follows the convention in C07 for the convenience of proper motion analysis. The counterpart for each component at different frequencies are mainly determined by their relative positions to the core.  A newly identified component at 8.4 and 22 GHz which lies at about 4 mas away from the northmost component is labeled as \emph{n}. This component was not detected in the 2001-2002 session 1 observations of C07, which may be due to its weakness and relative closeness to component \emph{d} during the observing campaign.

\subsection{Spectrum and identification of components}
\begin{figure}
   \includegraphics[scale=0.4]{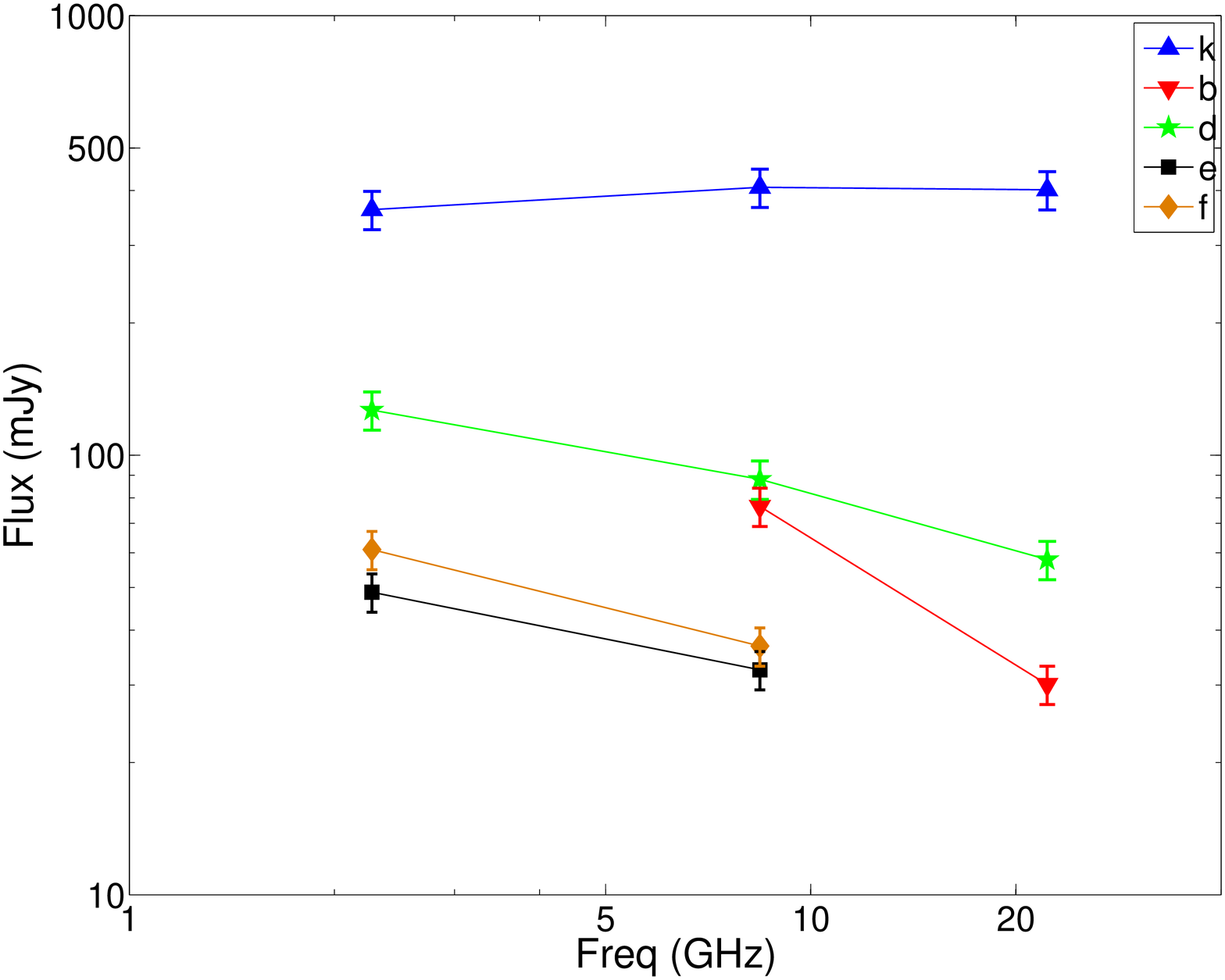}
   \includegraphics[scale=0.4]{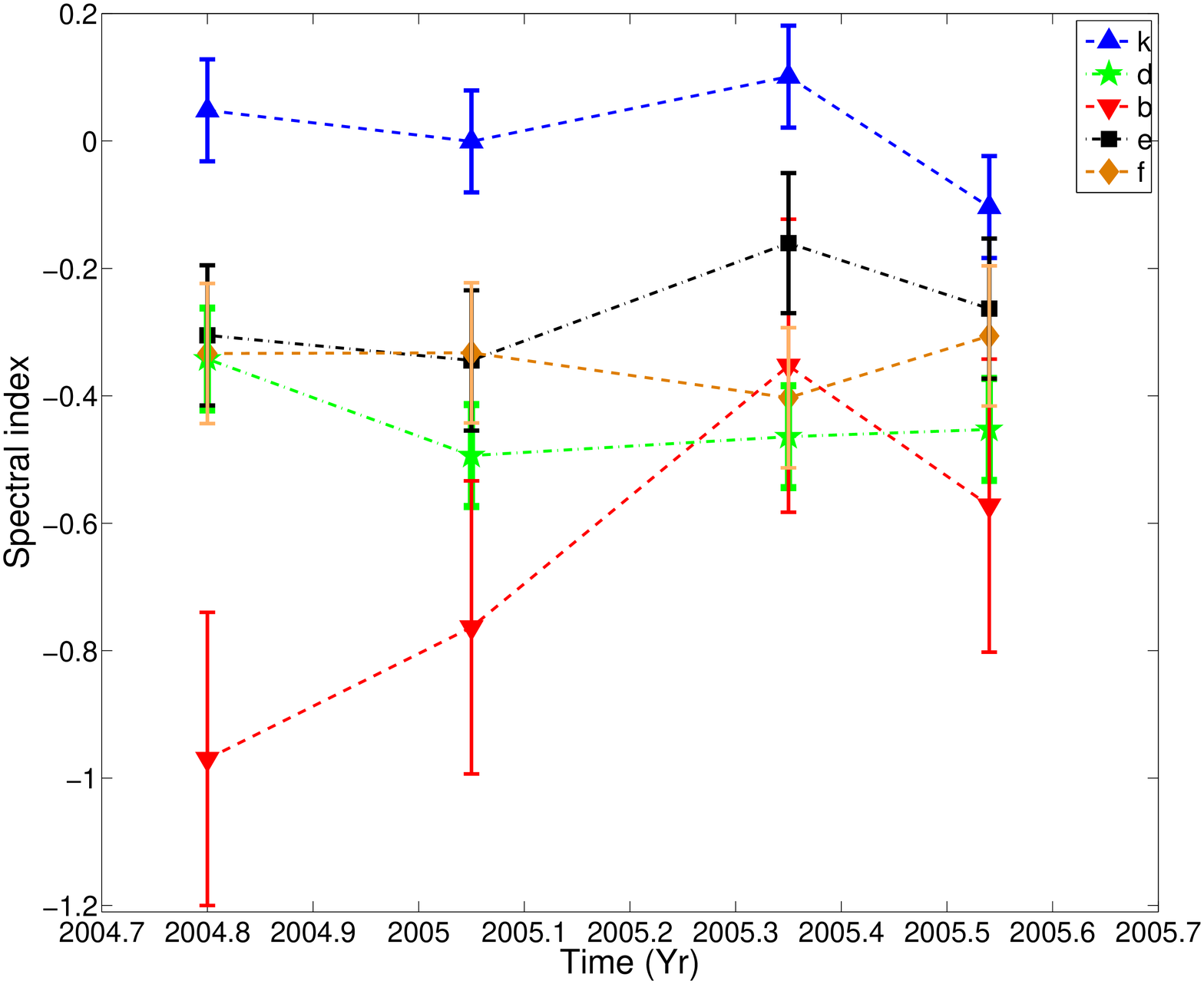}
   \caption{Top: Component spectra of 3C 66A on 2004.80. Bottom: spectral indices of jet components at all epochs}
   \label{fig:3}
   \end{figure}

The detection of components \emph{k}, \emph{b}, \emph{d}, \emph{e},
and \emph{f} at more than one frequency allows us to make spectral analysis of these components. Fig.~\ref{fig:3} (top) displays the spectra of these components at the first epoch (2004.80) based on the flux densities obtained by model-fitting of Gaussian components to the visibility data. The spectral indices ($S\propto\nu^\alpha$) at all epochs are shown in Fig.~\ref{fig:3} (bottom), among which the spectral indices
of components \emph{k} and \emph{d} are obtained through fits to the data at all
the 3 frequencies, and the others are obtained using data at 2
frequencies at which the components were detected. The component \emph{k} at the north end of the whole radio structure shows the flattest spectrum ($\alpha \sim 0$) among all these components, which enables us to identify this component as the core. By contrast, the other components show relatively steep spectra ($\alpha < 0$) .

We also estimate the brightness temperatures of all the components in
the source frame by using the expression (Shen et al. 1997)
\begin{equation}
{\normalsize T}_{\small b}~=~1.22\times10^{12}~{\normalsize S}~\nu^{-2}~
{\normalsize d}^{-2}~(1+{\normalsize z})~ {\normalsize K}~~,
\end{equation}
where S is the flux density in Jy, $\nu$ is the observational frequency in GHz, d is the diameter in mas, and \emph{z} is the redshift. The
brightness temperatures were shown in the last collum in
Table.~\ref{tab:2}. Component \emph{k} has the highest brightness temperature among all components, which greatly exceeds the Compton limit of $10^{12} K$ and hence gives further supports to our identification of the core component, as is shown in Table ~\ref{tab:2}.

\subsection{Proper motions of the jet components} \label{kinematic}
\begin{figure}
   \centering
   \label{fig:4a}%
   \includegraphics[scale=0.25]{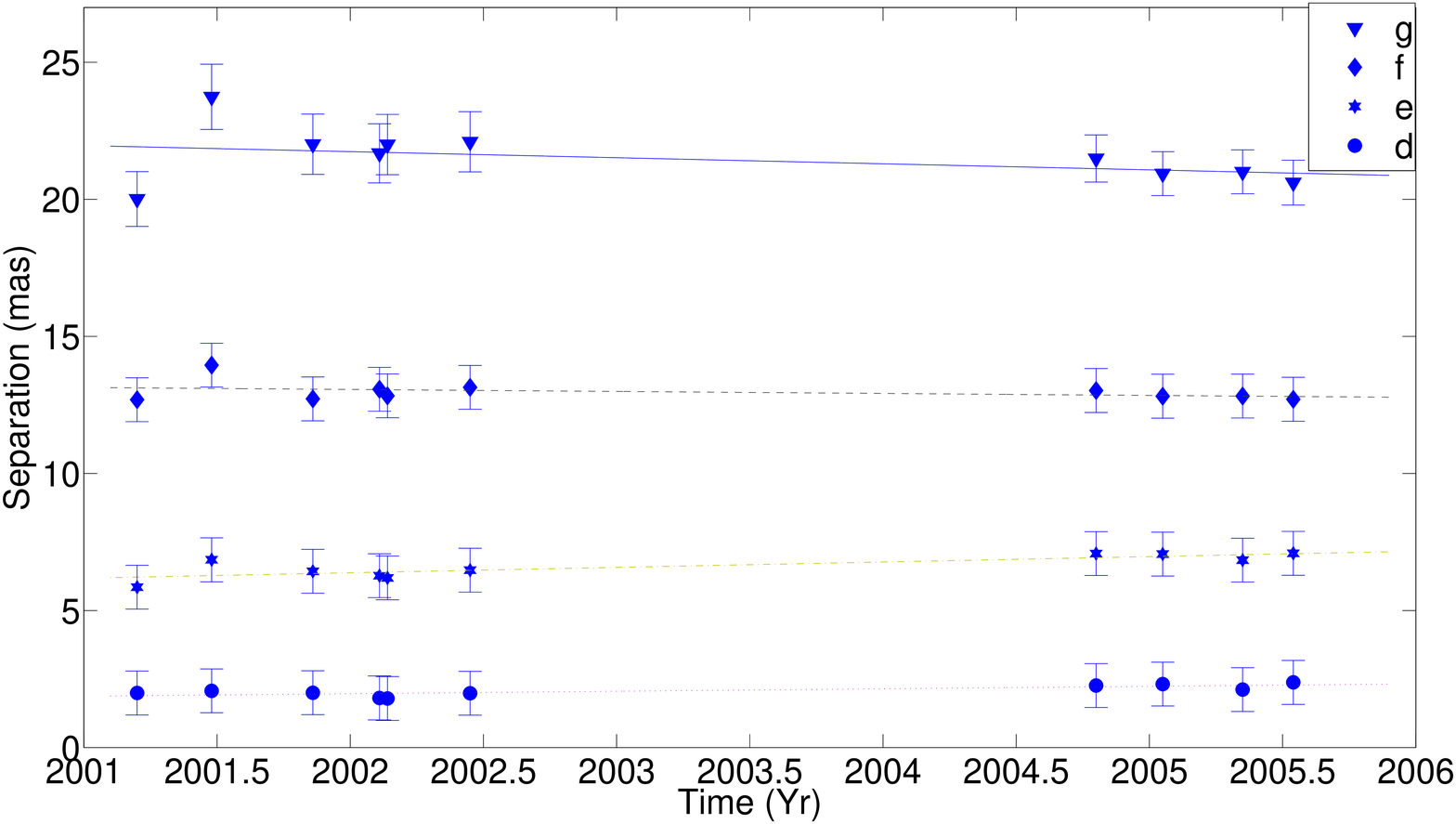}
   \label{fig:4b}%
   \includegraphics[scale=0.25]{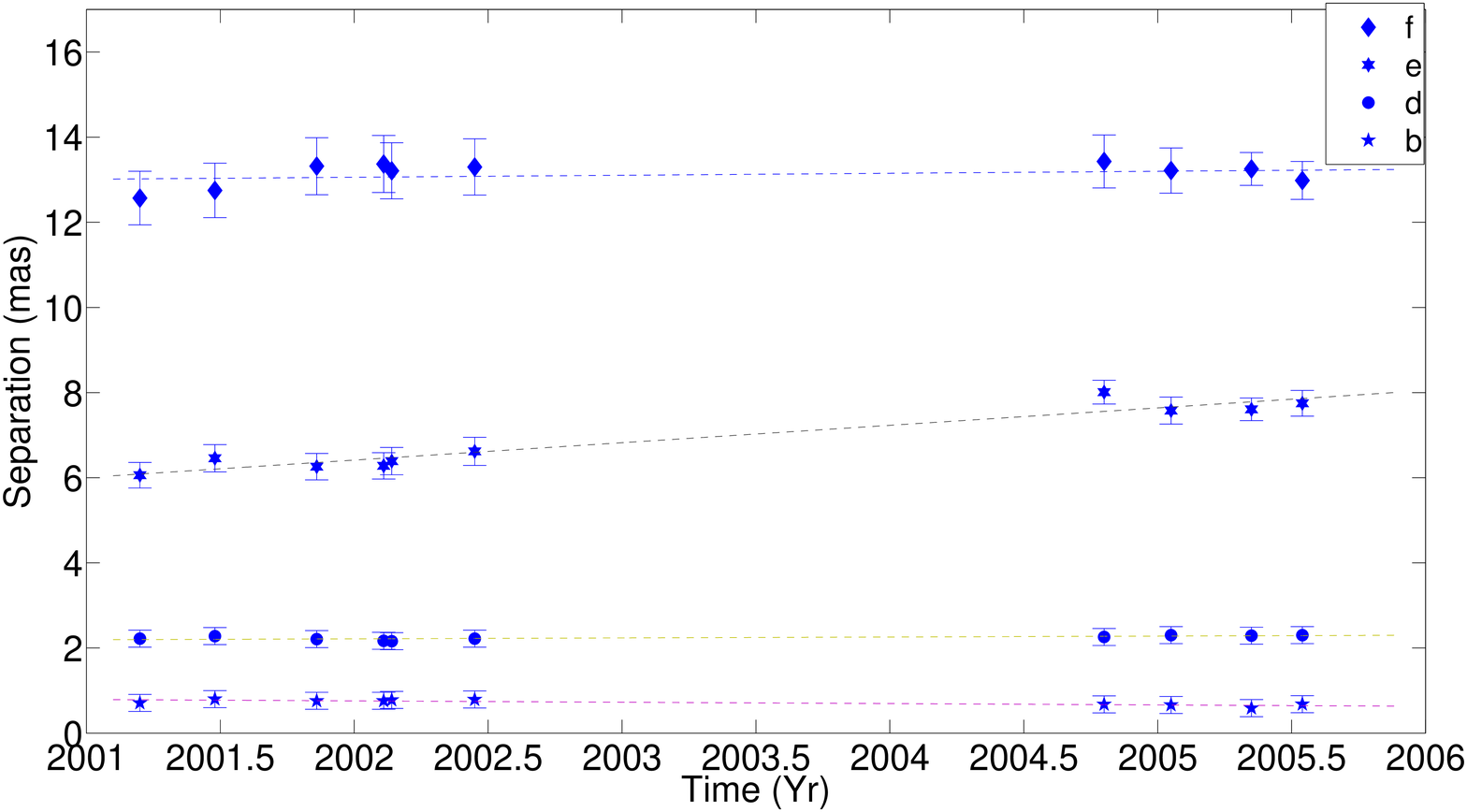}
   \label{fig:4c}%
   \includegraphics[scale=0.25]{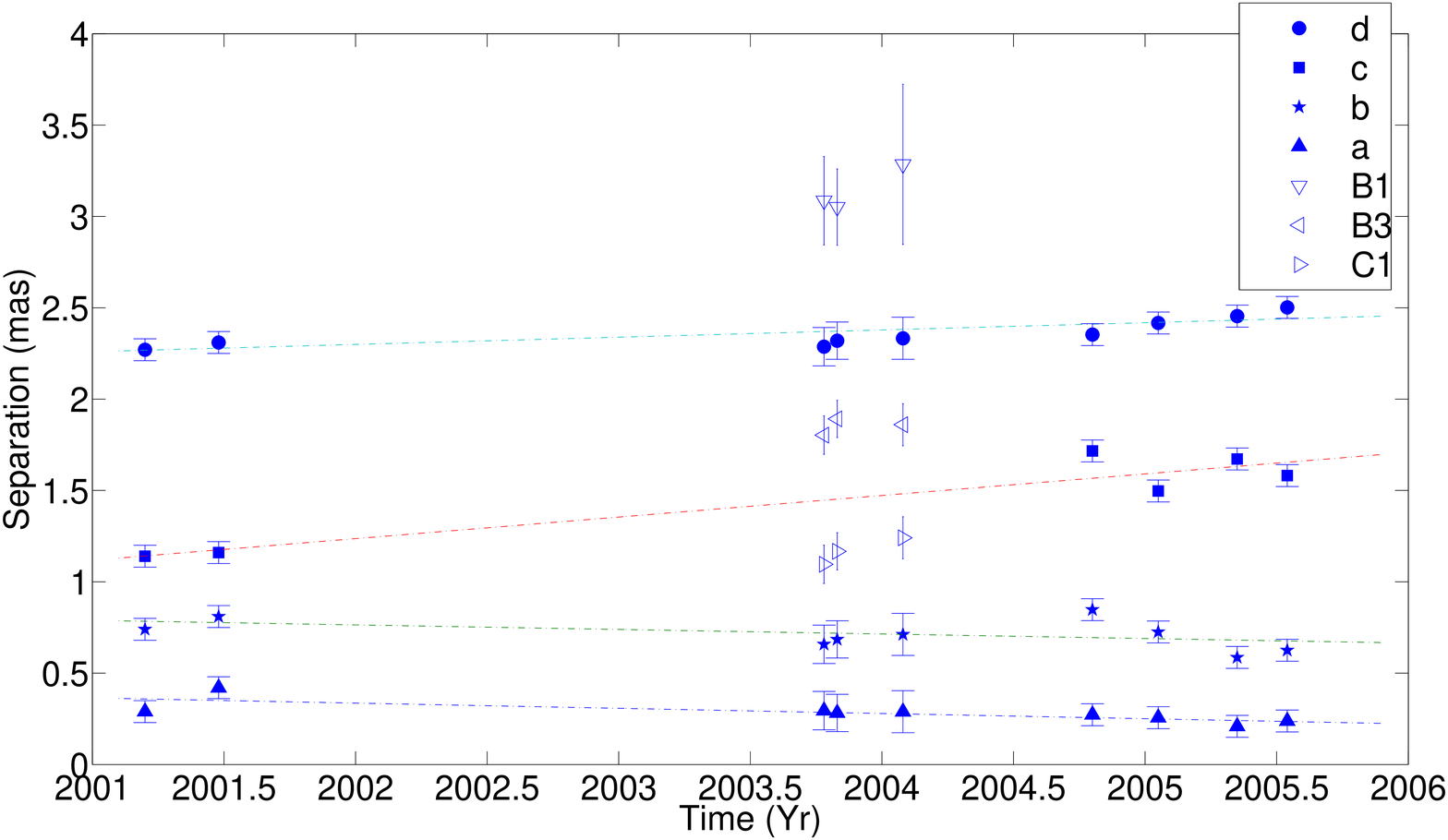}
   \caption{Separations from the core as functions of time for the jet components in 3C 66A at 2.3 GHz (top), 8.4 GHz (middle), and 22 GHz (bottom). The 2001-2002 results are taken from C07, 2003-2004 results at 22 GHz are taken from B05. The open symbols represents the components that cannot be cross identified.}
   \label{fig:4}
   \end{figure}

C07 studied the kinematics of the jet components in 3C 66A at 2.3
and 8.4 GHz using Session 1 data and no obvious proper motion was
detected within the uncertainties because of the short time coverage
($\sim$1.2yr). They combined their results with some previous works
(Jorstad et al. 2001, Jorstad et al. 2005, and
B05) and they found the kinematics of 3C 66A is rather complicated,
yet the observations were carried
out at different frequencies. By combining the data of the two sessions, we are able to study the kinematics of 3C 66A over a much longer time range (more than 4 years). The 22 GHz preliminary results were already published in the proceeding of IAU Symposium 290 (Zhao et al. 2013). In Fig.~\ref{fig:4}, we show the plots of separations of jet components from the core that have measurements at both sessions as a function of time. The apparent velocity obtained by linear fit to the separations are shown in Table.~\ref{tab:4}. We compared our 22 GHz results with that of B05 which was obtained in the time gap between our two Sessions (2003-2004). The model fitted positions of components \emph{a}, \emph{b}, and \emph{d} are in good agreement with C3, C2, and B2, respectively, in their work (Fig.~\ref{fig:4}c), while component \emph{c} seems to be divided into two components (B3 or C1). So we also included B05 results (Comp. C3, C2, \& B2) in our 22 GHz proper motion study.

 Our results show that the speeds of components in the jet of 3C 66A are much slower than previously reported (e.g. Jorstad et al. 2001, 2005). The results confirmed that the kinematics of this source is complicated, i.e. some components show superluminal motions, some show inward motions,
and the others are nearly stable. As shown in Fig.~\ref{fig:4}, there are no obvious proper motions
found for component \emph{f} at $\geq 1\sigma$ level. Components \emph{a}, \emph{b}, and \emph{g} show inward motions at $> 1\sigma$ level
while components \emph{c}, \emph{d}, and \emph{e} move downstream the jet superluminally at $\geq 3\sigma$ level.

\begin{deluxetable}{ccccccc}
\tablecolumns{7}
\tablewidth{0pc}
\tablecaption{Parameters of the Gaussian model fit components of the images shown in Fig.~\ref{fig:1}}
\tablehead{
\colhead{Freq. (GHz)}& \colhead{Comp.}& \colhead{S (Jy)}&\colhead{R (mas)}&
\colhead{$\theta$ (deg)}& \colhead{Major (mas)} &\colhead{Tb (K)}}
\startdata
\multicolumn{7}{l}{Epoch 2004.80}\\\hline
S&k &0.362$\pm$0.036&0& 0&0.35$\pm$0.00&$1.01\times10^{12}$\\
&d &0.127$\pm$0.013&2.26$\pm$0.80&-176.9$\pm$0.1&0.82$\pm$0.01&$6.28\times10^{10}$\\
&e &0.049$\pm$0.005&7.08$\pm$0.80&170.1$\pm$0.5&2.94$\pm$0.13&$1.90\times10^{09}$\\
&f &0.061$\pm$0.006&13.03$\pm$0.80&162.9$\pm$1.0&4.92$\pm$0.47&$8.45\times10^{08}$\\
&g &0.032$\pm$0.003&21.49$\pm$0.86&167.4$\pm$2.3&9.24$\pm$1.72&$1.23\times10^{08}$\\\hline

X&k&0.401$\pm$0.040&0&0&0.09$\pm$0.00&$1.11\times10^{12}$\\
&b&0.076$\pm$0.008&0.68$\pm$0.20&-171.7$\pm$0.6&0.30$\pm$0.01&$2.19\times10^{10}$\\
&d&0.088$\pm$0.009&2.26$\pm$0.20&-173.4$\pm$0.4&0.58$\pm$0.03&$6.66\times10^{09}$\\
&n&0.018$\pm$0.002&4.20$\pm$0.23&179.8$\pm$3.2&1.99$\pm$0.47&$1.15\times10^{08}$\\
&e&0.033$\pm$0.003&8.01$\pm$0.28&170.0$\pm$2.0&2.82$\pm$0.56&$1.02\times10^{08}$\\
&f&0.037$\pm$0.004&13.43$\pm$0.62&164.4$\pm$2.7&4.26$\pm$1.25&$5.06\times10^{07}$\\\hline

K&k&0.402$\pm$0.080&0&0&0.04$\pm$0.00&$9.02\times10^{11}$\\
&a&0.077$\pm$0.015&0.27$\pm$0.06&-171.4$\pm$0.5&0.13$\pm$0.00&$1.61\times10^{10}$\\
&b&0.030$\pm$0.006&0.85$\pm$0.06&-174.7$\pm$0.8&0.35$\pm$0.02&$8.63\times10^{08}$\\
&c&0.007$\pm$0.001&1.72$\pm$0.06&-166.0$\pm$1.3&0.33$\pm$0.08&$2.21\times10^{08}$\\
&d&0.058$\pm$0.012&2.35$\pm$0.06&-172.8$\pm$0.4&0.42$\pm$0.03&$1.19\times10^{09}$\\
&n&0.006$\pm$0.001&3.99$\pm$0.15&-174.0$\pm$2.2&0.67$\pm$0.31&$4.91\times10^{07}$\\\hline
\multicolumn{7}{l}{Epoch 2005.05}\\\hline
S&k &0.383$\pm$0.038&0& 0&0.13$\pm$0.00&$7.63\times10^{12}$\\
&d &0.130$\pm$0.013&2.32$\pm$0.80&-175.7$\pm$0.1&0.87$\pm$0.01&$5.65\times10^{10}$\\
&e &0.050$\pm$0.005&7.06$\pm$0.80&170.6$\pm$0.7&2.97$\pm$0.18&$1.87\times10^{09}$\\
&f &0.061$\pm$0.006&12.82$\pm$0.80&163.3$\pm$1.1&4.89$\pm$0.49&$8.51\times10^{08}$\\
&g &0.034$\pm$0.003&20.94$\pm$0.80&166.9$\pm$2.2&8.47$\pm$1.60&$1.57\times10^{08}$\\\hline

X&k&0.430$\pm$0.043&0&0&0.10$\pm$0.00&$1.05\times10^{12}$\\
&b&0.063$\pm$0.006&0.66$\pm$0.20&-174.1$\pm$0.6&0.32$\pm$0.01&$1.51\times10^{10}$\\
&d&0.076$\pm$0.008&2.30$\pm$0.20&-173.9$\pm$0.4&0.60$\pm$0.03&$5.17\times10^{09}$\\
&n&0.014$\pm$0.001&3.96$\pm$0.20&-178.5$\pm$2.5&1.65$\pm$0.35&$1.27\times10^{08}$\\
&e&0.032$\pm$0.003&7.58$\pm$0.32&169.2$\pm$2.4&2.87$\pm$0.64&$9.75\times10^{07}$\\
&f&0.039$\pm$0.004&13.22$\pm$0.53&163.4$\pm$2.3&4.16$\pm$1.06&$5.64\times10^{07}$\\\hline

K&k&0.378$\pm$0.076&0&0&0.06$\pm$0.00&$3.48\times10^{11}$\\
&a&0.078$\pm$0.016&0.26$\pm$0.06&-173.5$\pm$0.4&0.14$\pm$0.00&$1.38\times10^{10}$\\
&b&0.030$\pm$0.006&0.73$\pm$0.06&-174.8$\pm$0.6&0.27$\pm$0.01&$1.55\times10^{09}$\\
&c&0.006$\pm$0.001&1.50$\pm$0.06&-170.5$\pm$0.6&0.22$\pm$0.03&$4.41\times10^{08}$\\
&d&0.042$\pm$0.008&2.42$\pm$0.06&-173.0$\pm$0.7&0.47$\pm$0.06&$7.06\times10^{08}$\\
&n&0.007$\pm$0.001&3.67$\pm$0.31&-173.9$\pm$4.8&1.41$\pm$0.62&$1.29\times10^{07}$\\\hline
\multicolumn{7}{l}{Epoch 2005.35}\\\hline
S&k &0.317$\pm$0.032&0& 0&0.05$\pm$0.00&$4.18\times10^{13}$\\
&d &0.126$\pm$0.013&2.11$\pm$0.80&-175.1$\pm$0.0&0.76$\pm$0.00&$7.33\times10^{10}$\\
&e &0.051$\pm$0.005&6.84$\pm$0.80&170.9$\pm$0.5&2.89$\pm$0.13&$2.02\times10^{09}$\\
&f &0.062$\pm$0.006&12.82$\pm$0.80&163.1$\pm$1.1&4.75$\pm$0.48&$9.21\times10^{08}$\\
&g &0.034$\pm$0.003&21.00$\pm$0.80&166.5$\pm$2.2&9.02$\pm$1.58&$1.38\times10^{08}$\\\hline

X&k&0.427$\pm$0.043&0&0&0.14$\pm$0.00&$5.79\times10^{11}$\\
&b&0.093$\pm$0.009&0.59$\pm$0.20&-174.6$\pm$0.3&0.39$\pm$0.01&$1.55\times10^{10}$\\
&d&0.069$\pm$0.007&2.29$\pm$0.20&-174.2$\pm$0.2&0.54$\pm$0.02&$6.01\times10^{09}$\\
&n&0.013$\pm$0.001&3.48$\pm$0.20&-178.1$\pm$0.7&0.91$\pm$0.08&$3.89\times10^{08}$\\
&e&0.040$\pm$0.004&7.61$\pm$0.26&169.8$\pm$2.0&3.14$\pm$0.53&$1.00\times10^{08}$\\
&f&0.035$\pm$0.004&13.25$\pm$0.39&162.5$\pm$1.7&3.50$\pm$0.77&$7.19\times10^{07}$\\\hline

K&k&0.392$\pm$0.078&0&0&0.02$\pm$0.00&$5.94\times10^{12}$\\
&a&0.073$\pm$0.015&0.21$\pm$0.06&-163.2$\pm$0.2&0.03$\pm$0.00&$2.59\times10^{11}$\\
&b&0.066$\pm$0.013&0.59$\pm$0.06&-174.1$\pm$0.3&0.15$\pm$0.01&$1.09\times10^{10}$\\
&c&0.011$\pm$0.002&1.67$\pm$0.06&-171.5$\pm$0.1&0.14$\pm$0.01&$2.16\times10^{09}$\\
&d&0.044$\pm$0.009&2.45$\pm$0.06&-174.5$\pm$0.4&0.46$\pm$0.03&$7.67\times10^{08}$\\
&n&0.006$\pm$0.001&3.69$\pm$0.13&-179.4$\pm$2.1&0.88$\pm$0.27&$2.84\times10^{07}$\\\hline
\multicolumn{7}{l}{Epoch 2005.54}\\\hline
S&k &0.391$\pm$0.039&0& 0&0.30$\pm$0.00&$1.45\times10^{12}$\\
&d &0.105$\pm$0.011&2.38$\pm$0.80&-175.5$\pm$0.2&0.92$\pm$0.01&$4.14\times10^{10}$\\
&e &0.045$\pm$0.005&7.09$\pm$0.80&170.9$\pm$0.6&3.02$\pm$0.14&$1.65\times10^{09}$\\
&f &0.062$\pm$0.006&12.71$\pm$0.80&162.8$\pm$0.9&4.64$\pm$0.41&$9.63\times10^{08}$\\
&g &0.035$\pm$0.003&20.61$\pm$0.82&166.9$\pm$2.3&9.23$\pm$1.64&$1.35\times10^{08}$\\\hline

X&k&0.388$\pm$0.039&0&0&0.13$\pm$0.00&$5.40\times10^{11}$\\
&b&0.075$\pm$0.008&0.68$\pm$0.20&-176.3$\pm$0.5&0.24$\pm$0.01&$3.19\times10^{10}$\\
&d&0.062$\pm$0.006&2.30$\pm$0.20&-174.7$\pm$0.5&0.60$\pm$0.04&$4.35\times10^{09}$\\
&n&0.018$\pm$0.002&3.70$\pm$0.20&-177.9$\pm$2.7&1.51$\pm$0.35&$1.94\times10^{08}$\\
&e&0.032$\pm$0.003&7.75$\pm$0.30&170.5$\pm$2.2&2.85$\pm$0.60&$9.96\times10^{07}$\\
&f&0.037$\pm$0.004&12.98$\pm$0.45&161.8$\pm$2.0&3.53$\pm$0.89&$7.44\times10^{07}$\\\hline

K&k&0.306$\pm$0.061&0&0&0.04$\pm$0.00&$6.19\times10^{11}$\\
&a&0.094$\pm$0.019&0.24$\pm$0.06&-164.9$\pm$0.4&0.12$\pm$0.00&$2.31\times10^{10}$\\
&b&0.043$\pm$0.009&0.63$\pm$0.06&-175.1$\pm$0.5&0.21$\pm$0.01&$3.73\times10^{09}$\\
&c&0.008$\pm$0.002&1.58$\pm$0.06&-172.5$\pm$2.0&0.50$\pm$0.11&$1.21\times10^{08}$\\
&d&0.037$\pm$0.007&2.50$\pm$0.06&-173.9$\pm$0.5&0.45$\pm$0.05&$6.62\times10^{08}$\\
&n&0.004$\pm$0.001&3.36$\pm$0.06&-179.5$\pm$0.3&0.13$\pm$0.04&$8.45\times10^{08}$\\
\enddata

\label{tab:2}
\end{deluxetable}

%
%

\begin{table*}
\begin{center}\caption{The proper motions of the jet components in 3C~66A} \label{tab:4} \vspace{2mm}
\small
\begin{tabular}{ccccccc}\hline\hline
&\multicolumn{2}{c}{2.3 GHz}&\multicolumn{2}{c}{8.4
GHz}&\multicolumn{2}{c}{22.2 GHz}\\ &$\mu$
&$\beta_{app}(z=0.444)$ &$\mu$ &$\beta_{app}(z=0.444)$ &$\mu$ &$\beta_{app}(z=0.444)$\\
 & ${\rm mas\,yr}^{-1}$ & & ${\rm mas\,yr}^{-1}$ & & ${\rm mas\,yr}^{-1}$ & \\\hline
a & & & & &-0.028$\pm$0.013 &-0.76$\pm$0.36 \\
b & & &-0.032$\pm$0.038 &-0.85$\pm$1.02 &-0.025$\pm$0.013 &-0.67$\pm$0.36 \\
c & & & & &0.118$\pm$0.013 &3.16$\pm$0.36 \\
d &0.089$\pm$0.152 &2.38$\pm$4.07 &0.021$\pm$0.038 &0.57$\pm$1.02 &0.040$\pm$0.013 &1.06$\pm$0.36 \\
e &0.194$\pm$0.152 &5.24$\pm$4.07 &0.409$\pm$0.057 &10.94$\pm$1.52 & & \\
f &-0.072$\pm$0.152 &-1.93$\pm$4.07 &0.047$\pm$0.103 &1.25$\pm$2.75 & & \\
g &-0.222$\pm$0.177 &-5.94$\pm$4.74 & & & & \\ \hline

\end{tabular}
\end{center}
Notes: $\beta_{app}(z=0.444)$ is the corresponding apparent speed for $\mu$ at z=0.444. Considering the redshift range of 3C 66A ($0.335 < z < 0.444$), $\beta_{app}$ should be $0.78\beta_{app}(z=0.444) < \beta_{app} < \beta_{app}(z=0.444)$
\end{table*}

\section{Discussion}
\label{sect:discussion}

\subsection{Morphology}

 Our VLBI images of 3C 66A show similar
morphology as previous works, which is a typical one-side jet
with two bendings at about 1.2 and 4 mas from the core. Jet bendings or non-linear morphologies are very common in BL Lacs (e.g. Karouzos et al. 2012). Jorstad et al. (2005) suggest that the first bending in 3C 66A jet occurs smoothly from 1.2 mas up to 3 mas. The former works mentioned above all show that the flux densities of the components at 22 and
43 GHz decrease along the jet but start to increase at 1.2 mas up to
2.6 mas. Such a coincidence suggests that the observed bending may be
related to the cause of the local flux density increase in the jet, which points directly to the continues decrease of the viewing angles
and hence the increase of Doppler boosting effect. The model fitting
results of components k-d in our 22 GHz maps also support this
argument. Bendings could also be originated by interaction with ambient medium (e.g. Alberdi et al. 2000), but this is less likely to be the case for the first bending in 3C 66A because there are no stationary components in this area (see Sec.~\ref{kinematic}).

As measured in the other blazar objects, the detectable radio structure of 3C 66A gets smaller with frequency.
The restored emission region at 22 GHz is almost 6 times smaller than
that at 2.3 GHz as a result of the much higher resolution plus the decreased surface brightness at
22 GHz. Note that the brightness temperature drops with
frequency, indicating its decreased surface brightness at
higher frequency, as is suggested in Kellermann et al. (2004).

\subsection{Kinematics}
 We fit the core separations of the jet components of 3C 66A for the
two Sessions, which indicates the kinematics of 3C 66A is very
complicated. We have detected apparent inward motions as well as superluminal motions. The superluminal moving components (\emph{c}, \emph{d}, \emph{e}) are found between 1.5 and 8 mas from the core. Following Lu et al.~(2011), we could put constrains on basic jet parameters using the apparent speed, e.g. $\Gamma_{min}=\sqrt{1+\beta^{2}_{app}}$ and $\theta_{max}=2\arctan(1/\beta_{app})$. For the fastest component, \emph{e} ($8.49 < \beta_{app} < 10.94$), we have, $8.55 < \Gamma_{min} < 10.99$, $10.4^{\circ} < \theta_{max} < 13.4^{\circ}$, indicating that the Doppler boosting effect may play quite a large role in the very high core brightness temperature and detection of $\gamma$-ray emission.

\begin{figure}
   \centering
   \includegraphics[scale=0.35]{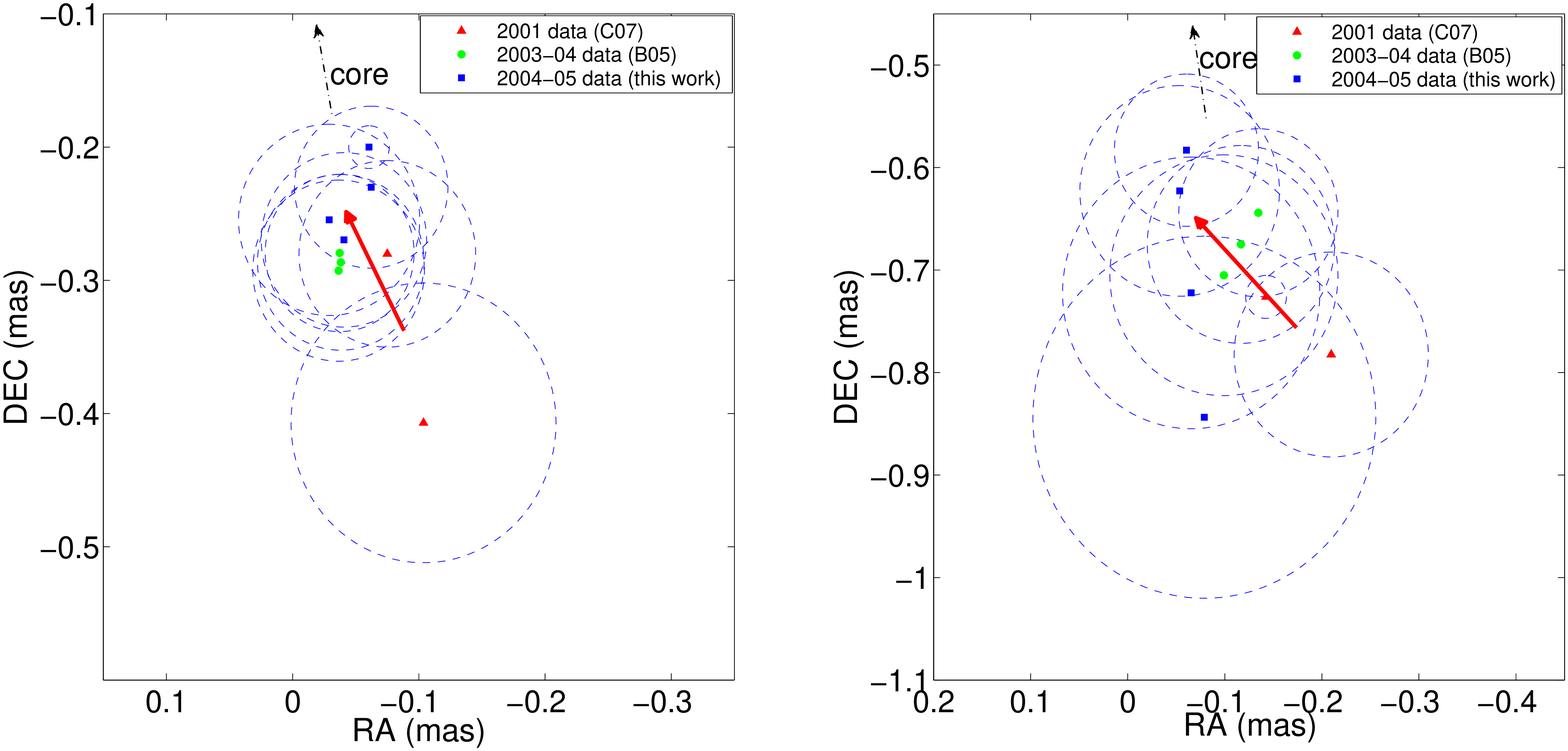}
   \caption{2D position plot of Component a (left) and b (right) at 22 GHz. The triangles, circles, and squares represent the results from C07, B05, and this work, respectively. The arrow shows the average trend of the component motion. (0,0) represents the position of the core.}
   \label{fig:2dp}
   \end{figure}

Apparent inward motions have been detected in a number of individual objects (e.g. Mrk 421, Niinuma et al. 2012, and OJ 287, Sawada-Satoh et al. 2013). But large sample studies have shown that such motions are relatively rare (Kellermann et al. 2004, Lister et al. 2013). Lister et al. (2013) reported that 17 out of 887 components ($< 2\%$) show inward motions and that most of inward motions are slow ($< 100~\mu as~yr^{-1}$) and occur within $\sim$ 1 mas from the core. We have detected inward motions in both the innermost (comp. \emph{a} and \emph{b}) and outermost region (comp. \emph{g}) of the jet in 3C 66A. The motion of component g is probably due to the change of internal brightness distribution in this feature because it is relatively weak and extended. The reason for inward motions of the innermost components, as suggested by previous studies (e.g. Kellermann et al. 2004), could be: (1) the motion of a newly emerged component which is still very close to the core and could not be separated by current resolution; (2) highly curved jet motions that cross the line of sight (LOS). The inward motions of the innermost components in 3C 66A were first reported by us at $> 2\sigma$ confidence (Zhao et al. 2013) and were then confirmed by the MOJAVE results at 15 GHz (Lister et al. 2013). The MOJAVE data cover mainly the time after our observation and we can cross identify most of our components with the 15 GHz ones (a, b, c, e, and f in our result with 13, 11, 9, 2, and 1 in MOJAVE results) by the extrapolation of the fitted motions. This means despite of a system shift of the component positions, which is caused by frequency-dependent core positions (i.e. the core-shift effect), the components at the inner part of the jet is continuously showing apparent-inward motions for $\sim$ 10 years (2001 to 2011) and no new components were detected in this area even during and after $\gamma$-ray and radio flares (e.g. in October 2008, Abdo et al. 2011). This result rules out both the two above explanations because otherwise for the former, the new-born component would show up after such a long time and for the later, the highly curved motions should turn over and cross the LOS again so the motions would change to outward in 10 years. Here we propose another possible explanation: non-stationarity of the core due to opacity change. The non-stationarity of the core has been detected for some sources (e.g Mrk 421, Niinuma et al. in preparation, and 3C 454.3, Bartel et al. 2012) and the opacity change in 3C 66A is supported by the 2 epoch core-shift measurements in Pushkarev et al. (2012). Fig~\ref{fig:2dp} shows the 2D position plot of components \emph{a} and \emph{b} at 22 GHz. The average trend for the two components (bold arrow) are similar, which also indicate the change of the core position.
The reason such a trend is not obvious in the other components is that these outer components were ejected much earlier than the inner components and according to the previous works the reported speeds were much faster (Jorstad et al. 2001, 2005). So their own motions dominate the overall motion. While for the innermost components, they are slower, even stable, so the core motion dominates the observed relative motions.
Such a similarity could also be found in the MOJAVE results of this source. Note that the 3 third component in MOJAVE results is also likely inward moving, while it shows outward motions in 2001-2005 (Figure 3). This could be explained as this component decelerated around 2005 and began to show
a similar motion as the inner 2 components.

Future precise measurements of the core-shift with multi-frequency astrometric observations
and detailed analysis of the variability of the core-shift and the possible correlation with
flux, spectral variabilities would be important to test this explanation.



\section{Summary}

\label{sect:sum} We studied the parsec-scale jet properties
of the TEV blazar 3C 66A by using multi-epoch VLBI observations at 2.3, 8.4, and 22
GHz from 2004 to 2005. Our maps of 3C 66A show similar morphology as previous works, i.e. a core-jet structure with bendings.
The northmost component, \emph{k} is identified as the core based on its flat
spectrum and high brightness. The other components are found to be weaker and show steeper spectra.

We studied the proper motion of jet components in 3C 66A over a
timerange of more than 4 years and we found the kinematics of this
source is very complicated. We detected superluminal motions
for 3 components. We found the innermost components show inward motions. By combining with the results at 15 GHz, we have ruled out the possibility of new-born components or highly curved jet trajectories as reasons of the inward motion. We argue that the possible reason could be non-stationarity of the core. Further observations are needed to unveil the nature of this phenomenon.

\begin{acknowledgements}

\textbf{This work is partly supported by National Basic Research Program of China (973 program) No. 2012CB821806, the National Natural Science Foundation of China (Grants 10625314, 11121062, and 11273042), the CAS/SAFEA International Partnership Program for Creative Research Teams, the Strategic Priority Research Program on Space Science, the Chinese Academy of Sciences (Grant No. XDA04060700), and the Science and Technology Commission of Shanghai Municipality
(Grant No. 12ZR1436100).}

\end{acknowledgements}

\end{document}